\documentstyle[preprint,aps]{revtex}
\newcommand{\be}{\begin{eqnarray}}
\newcommand{\ee}{\end{eqnarray}}
\newcommand{\ra}{\rightarrow}
\begin{document}
\draft
\preprint{KAIST--CHEP--95/03}
\vspace{4.0in}
\title{ Approximate Flavour Symmetries\\
        and See-Saw Mechanism }
\vspace{1.0in}
\author{
Kang Young Lee\thanks{kylee@chep5.kaist.ac.kr}
and Jae Kwan Kim
}
\vspace{.5in}
\address{
Dept. of Physics, KAIST, Taejon 305-701, KOREA \\
}
\date{\today}
\maketitle
\begin{abstract}
\\
We study the approximate flavour symmetries imposed on
the lepton sector assuming see-saw mechanism
as the neutrino mass structure.
We apply the symmetry to various neutrino phenomenologies
and obtain constraints on neutrino masses and mixings.
\end{abstract}
\pacs{ }

\narrowtext

\section{Introduction}

In the standard model (SM), there exists one Higgs doublet
as the only source of the electroweak symmetry breaking.
However many extensions of the SM need more complex
Higgs structure. When we suggest additional Higgs bosons,
we face some constraints. One of the major issues
with respect to multi-Higgs models is suppressions
of flavour-changing neutral currents (FCNC) mediated by the
exchange of neutral Higgs bosons.
Strong limits on FCNC have been given from several experiments such as
low rates of $K_L \ra \mu^+ \mu^-$, limits for
$K^+ \ra \pi^+ \nu {\bar \nu}$,
$D^0 \ra \mu^+ \mu^-$,
$B^0 \ra \mu^+ \mu^-$ processes and
$K^0-{\bar K}^0$, $B^0-{\bar B}^0$ mixings.

Following Glashaw-Weinberg criteria
one usually propose to impose a discrete symmetry on
the multi-Higgs models under which no Higgs doublet
couples to both u-type and d-type quark sectors.
This restriction is not inevitable, however,
we can give up the discrete symmetry and impose
flavour symmetries on fermions to suppress FCNC instead.
Without Yukawa interaction terms, the SM lagrangian manifests
a set of global U(1) flavour symmetries.
When we switch on Yukawa couplings, the flavour symmetries are
weakly broken and we have approximate symmetries because of
smallness of Yukawa couplings.
Recently it is proposed that the violations of the flavour
symmetries are parametrized by a set of small parameters
\{$\epsilon$\} \cite{ant,hall,rasin}.
It turns out that proper choice of the
small parameters can give suppressions of FCNC.

Hall and Weinberg \cite{hall} provide the analysis with particular
formulars for the small parameters on quark sectors
from observed masses and mixing angles.
They impose the suppression factors $\epsilon_{Qi}$
for the left-handed quark doublets, $\epsilon_{ui}$
and $\epsilon_{di}$ for the up-type and
down-type right-handed quarks respectively.
Then the Yukawa coupling constant $\lambda_{ijn}^U$ is
doubly suppressed by both $\epsilon_{i}$ and
$\epsilon_{j}$ such that the Yukawa couplings are of order
$\lambda_{ijn}^U \approx \epsilon_{Q_i}\epsilon_{u_j}$ and
$\lambda_{ijn}^D \approx \epsilon_{Q_i}\epsilon_{d_j}$
where $n$ is the index related to Higgs bosons.
They also show that CP must also be a good approximate symmetry
in their framework.
The problem of CP violation from multi-Higgs potential
with the approximate flavour symmetries is also pointed out
in the recent work by Wu and Wolfenstein \cite{wu}.

On the lepton sector, it is hard to specify the leptonic
suppression factors explicitly due to
ignorance of lepton mixing angles and neutrino masses.
Assuming neutrino masses with the famous
see-saw mechanism \cite{ss}, some statements can be made.
Ra$\check{{\rm s}}$in and Silva \cite{rasin} studied
the approximate flavour symmetries in the lepton sector
under the see-saw mechanism
and gave some predictions of neutrino masses and mixings
with additional Ans$\ddot{\mbox a}$tze.

In this letter we study the approximate flavour symmetries
on lepton sectors with the see-saw mchanism in detail and
apply them to several neutrino phenomenologies.
We show that the possible windows
for the neutrino masses and mixings
obtained by the MSW solution for the solar neutrino problem
can give some constraints on the neutrino
mass under the assumption of the approximate flavour symmetries.
We also discuss the possible existence of the neutrinoless
double beta decays, $(\beta \beta)_{0 \nu}$ and possibilities
to find the neutrino oscillation in CHORUS.

\section{See-Saw Mechanism with Approximate Flavour Symmetries}

We consider the neutrino mass terms with see-saw mechanism.
After the spontaneous symmetry breaking
the mass terms of neutrinos are given by
\begin{equation}
{\cal L}_{\nu} = -\frac{1}{2} {\bar \nu}^c_{Li} M^L_{ij} \nu_{Lj}
                 -\frac{1}{2} {\bar \nu}_{Ri} M^R_{ij} \nu^c_{Rj}
                 - {\bar \nu}_{Ri} M^D_{ij} \nu_{Lj}
\end{equation}
where the mass matrix $M^L_{ij}$ is of the left-handed Majorana neutrinos,
$M^R_{ij}$ of the right-handed Majorana neutrinos and $M^d_{ij}$ of the
Dirac neutrnos.
Let us assume that $M^L \sim 0$, $M^D \sim M_{fermions}$ and
$M^R \gg M_{fermions}$.
We write it as the compact form
\begin{equation}
{\cal L}_{\nu} = - \frac{1}{2}  \left( {\bar \nu^c_L}, {\bar \nu_R}
                                \right) M
                               \left( \begin{array}{c}
                                \nu_L \\
                                \nu^c_R
                               \end{array}   \right)
\end{equation}
where
\begin{equation}
M = \left( \begin{array}{c}
             ~~~0~~~~~M^D \\
             {M^D}^{\dagger}~~M^R
           \end{array}   \right)
  \equiv \left( \begin{array}{c}
             ~~~0~~~~~m \\
             {m}^{\dagger}~~M^R
           \end{array}   \right)
\end{equation}
We impose the suppression factors $\epsilon_{Li}$
for the left-handed lepton doublets and $\epsilon_{li}$ for the
right-handed charged leptons with respect to the ref. \cite{hall}.
The orderings $\epsilon_i \le \epsilon_j$ for $i \le j$
are assumed.
Let us impose the factor $\epsilon_{\nu_i}$ for right-handed
neutrinos. Since the suppression factors for the left-handed
neutrinos $\nu_L$ are common with the
charged lepton partners, we can assume
that $\epsilon^c_{Li} = \epsilon_{Li}$.
We also let $\epsilon^c_{Ri} = \epsilon_{Ri}$
with no loss of generality.
Then the mass matrices elements are written by
\be
M^D_{ij} \approx \epsilon_{Li} \epsilon_{\nu j}
                        \langle \varphi \rangle_1, ~~~~~
M^R_{ij} \approx \epsilon_{\nu i} \epsilon_{\nu j}
                        \langle \varphi \rangle_2.
\nonumber
\ee

Now we diagonalize the 6 $\times$ 6 mass matrix (3).
After block-diagonalization, we obtain
the matrix $M^{(bd)}$
\begin{eqnarray}
M^{(bd)} \sim \left( \begin{array}{c}
             -m{M^R}^{-1}m~~~~0 \\
             ~~~~~~0~~~~~~~~~~M^R
           \end{array}   \right)
           + {\cal O} \left( m \left( \frac{m}{M^R} \right)^2 \right)
\end{eqnarray}
where ${\cal O} (  m(m/M)^2 )$
describes matrices of which elements have the values
of order of magnitude of the square ratio
of the typical Dirac mass scale
to the typical right-handed Majorana mass scale.
Thus the block matrix elements are written as
\begin{eqnarray}
(m{M^R}^{-1}m)_{ij} &\sim& \epsilon_{Li} \epsilon_{Lj}
                \frac{\langle \varphi \rangle_1^2}
                     {\langle \varphi \rangle_2}
                \nonumber \\
M^R_{ij}          &\sim& \epsilon_{\nu i} \epsilon_{\nu j}
                     {\langle \varphi \rangle_2}.
\end{eqnarray}

Now we diagonalize the 3 $\times$ 3 matrices,
$ (m{M^R}^{-1}m)_{ij} $ and $ M^R_{ij} $ which are left- and
right-handed Majorana mass matrices.
In fact, we are interested in the left-handed Majorana
mass matrix and we obtain the relations as
\be
m_i &\approx& \epsilon_{Li}^2 \langle \varphi \rangle
       \nonumber \\
V_{ij}^{lep} &\approx& \epsilon_{Li}/\epsilon_{Lj}~~~~~~~~i,j = 1,2,3.
\ee
Therefore we get the simple relations
between masses and mixings for the neutrinos.
\be
V^{lep}_{ij} \approx \frac{\epsilon_{Li}}{\epsilon_{Lj}}
             \approx \sqrt{\frac{m_i}{m_j}}
\ee
Note that this is not a precise numerical calculations
but only order of magnitude estimates.

\section{MSW solutions and Approximate Flavor Symmetries}

The detection of neutrinos from the sun
is a possible probe to neutrino mixing hypothesis.
These experiments are sensitive to values of neutrino
masses and mixing angles so small
that could not be reached in laboratory experiments.
Ever since Davis \cite{dav}, several results are reported from
Kamiokande II \cite{kam}, SAGE \cite{sage} and GALLEX \cite{gallex}
collaborations.
Mikheyev and Smirnov \cite{ms} and Wolfenstein \cite{w}
have proposed a solution to the solar neutrino problem
from the point of view of resonance transitions of
neutrinos in matter.
All existing solar neutrino experimental data could be
described by the MSW mechanism
if we assume the standard solar model is valid.
The recent analysis found the following three windows
for the allowed values of the parameters
$\Delta m^2$ and sin$^2 2\theta$ \cite{pet1}.
\be
{\rm I.}~~~~~~~~~~~~~~~~~~3.2\times10^{-6}{\rm eV}^2
             \leq &\Delta m^2& \leq 1.2\times10^{-5}{\rm eV}^2
             \nonumber \\
      5.0\times10^{-3} \leq &\sin^2 2\theta& \leq 1.6\times10^{-2} \\
{\rm II.}~~~~~~~~~~~~~~~~~~5.4\times10^{-6}{\rm eV}^2
             \leq &\Delta m^2& \leq 1.1\times10^{-4}{\rm eV}^2
             \nonumber \\
         0.18 \leq &\sin^2 2\theta& \leq 0.86 \\
{\rm III.}~~~~~~~~~~~~~~~~~~1.0\times10^{-7}{\rm eV}^2
             \leq &\Delta m^2& \leq 1.8\times10^{-6}{\rm eV}^2
             \nonumber \\
             0.74 \leq &\sin^2 2\theta& \leq 0.93
\ee

We look into our result in above windows.
Following relation is derived from eq. (7)
\begin{equation}
\sin^2 2\theta = \frac{4}
                      {\sqrt{1+\frac{\Delta m^2}{m_1^2}}}
                \left( 1-\frac{1}{\sqrt{1+\frac{\Delta m^2}{m_1^2}}}
                \right)
\end{equation}
$m_1$ is the mass of the electron neutrino
and plays the role of an input parameter.
Keep in mind that the eq. (7) is not the exact one but only
the order of magnitude estimate and the relation (11) is
also the case.
Thus we cannot expect to extract the exact value of $m_1$
with the relation (11).
But we may estimate the order of magnitude of the mass
of the electron neutrino and it is sufficient at present
since we do not know even whether $\nu_e$ has non-zero
mass or not.
The results are shown in Fig. 1 for the different values of $m_1$.
We find that:
{\it i}) Window I favors the value $m_1 \sim {\cal O}(0.01)$ eV.
{\it ii}) Window II is matched  with the values of $m_1 < 0.1$
eV.
{\it iii})Window III favors small values of $m_1$, $m_1 < 0.01$.

\section{Possibilities of Neutrinoless Double-Beta Decays}

Searches of the neutrinoless double-beta decays ($(\beta \beta)_{0 \nu}$)
are known to be sensitive to the existence of massive Majorana
neutrinos coupled to the electron in the weak
charged lepton currents.
Since neutrino masses are expected to be small,
$m_{\nu} << 30$ MeV, the decay amplitude is proportional to
the (1,1) element of the Majorana mass matrix of neutrinos, $m_{ee}$.
We have the upper bound for $m_{ee}$, $m_{ee} < (1-2)$ eV from
the results of $(\beta \beta)_{0 \nu}$ decay searches.
Future experiments will cover the range of values of
$|m_{ee}|$ as small as $|m_{ee}| \simeq (0.1-0.3)$ eV.

Petkov and Smirnov \cite{pet2} studied the possibilities to
observe the $(\beta \beta)_{0 \nu}$ decay in the present and future
experiments with the windows obtained by the solar neutrino
experiments.
If one assume that the Majorana neutrinos obey the mass hierarchy
relation, $m_1 << m_2 << m_3$, the value of $|m_{ee}|$ can only
due to a sufficiently large admixture of the $\nu_3$ state with
a mass $m_3 \geq 0.1$ eV in the $\nu_e$ state:
$|m_{ee}| \sim m_3 |U_{e3}|$.
In this case, one have the mass hierarchy $m_1/m_3 \leq 10^{-3}$
and the large mixing between the first and the third families,
$|V_{13}| \leq 1$. This result is not compatible with the
eq. (7) from approximate flavour symmetries.

The case of highly degenerate neutrino mass spectrum is also
considered, $m_1 \simeq m_2 \simeq m_3 \geq 0.1$ eV.
The solar neutrino defict is then explained
by $\nu_e$ conversion to $\nu_{\mu}$ if
$|m_2-m_1| \leq (0.5-5)\times 10^{-5}$ eV.
And the atmospheric neutrino defict by the oscillation
$\nu_{\mu} \leftrightarrow \nu_{\tau}$ can also be
explained with $|m_3 - m_2| \simeq (0.5-5) \times 10^(-2)$ eV.
With these conditions eq. (7) requires large mixings
$|V_{ij}| \sim 1$ and the solar neutrino
windows II and III are permitted.
In the fig. 1, we find that an appreciable region may be allowed
around the window II which is compatible with the approximate flavour
symmetries with $m_1 \sim (0.1-0.01)$ eV
even though the value of $|m_{ee}|$ is somewhat small for
observation of $(\beta \beta)_{0 \nu}$ decay.
We conclude that under the assumption of the
approximate flavour symmetries on the
lepton sector observations of the $(\beta \beta)_{0 \nu}$ decay
imply the degenerate mass spectrum of neutrinos with the
mass $m_{\nu} < 0.1$ eV.

\section{Searches of $\nu_{\mu} \leftrightarrow \nu_{\tau}$
         oscillations}

Direct searches of neutrino oscillations at accelerator
proceed. CHORUS \cite{chorus} shall perform the experiment
to explore the domain of small mixing angles down to
$\sin^2 2 \theta_{\mu \tau} \sim 3 \times 10^{-4}$
for mass parameters $\Delta m^2 > 1$ eV$^2$.

Like eq. (11), we have the following relation for
$\nu_{\mu} \leftrightarrow \nu_{\tau}$ mixing.
\be
\sin^2 2\theta = 4 \sqrt{1-\frac{\Delta m^2}{m_3^2}}
                \left( 1-\sqrt{1-\frac{\Delta m^2}{m_3^2}}
                \right)
\ee
where $m_3$ is the tau neutrino mass.
We show that this relation for different values of $m_3$ with
the parameter space which will be covered by CHORUS experiment.
We find that we can estimate the tau neutrino mass $m_3$
under the assumption of the approximate flavour symmetries
if CHORUS observe the oscillation events in their experiment.

\section{Summary}

We studied the see-saw mechanism under the assumption of approximate
flavor symmetry.
This approximate symmetry provide guidelines for various
neutrino phenomenologies.
We derived a simple relation between neutrino masses
and mixing angles.
Using the relation, we gave several predictions for
the neutrino masses and mixings without additional
Ans$\ddot{\mbox a}$tze.
We showed that the relation is consistent with
MSW solution of the solar neutrino problem and
gave some constraints on the electron neutrino mass.
The neutrinoless double-beta decay seems not to be
compatible with the flavour symmetry well but
it would give strong conditions on neutrino masses and mixings
with the approximate flavour symmetry
if $(\beta \beta)_{0 \nu}$ decay is observed.
We also have another condition for the tau neutrino mass
if the neutrino oscillation is observed in CHORUS.

Though our calculations given here are not exact,
we can provide many useful estimates for the neutrino
masses and mixings which are tested in the near future.
\\
\\

\acknowledgements

We were supported in part by Korea Science and Engineering
Foundation (KOSEF).

%
%
%
\newpage

%
%

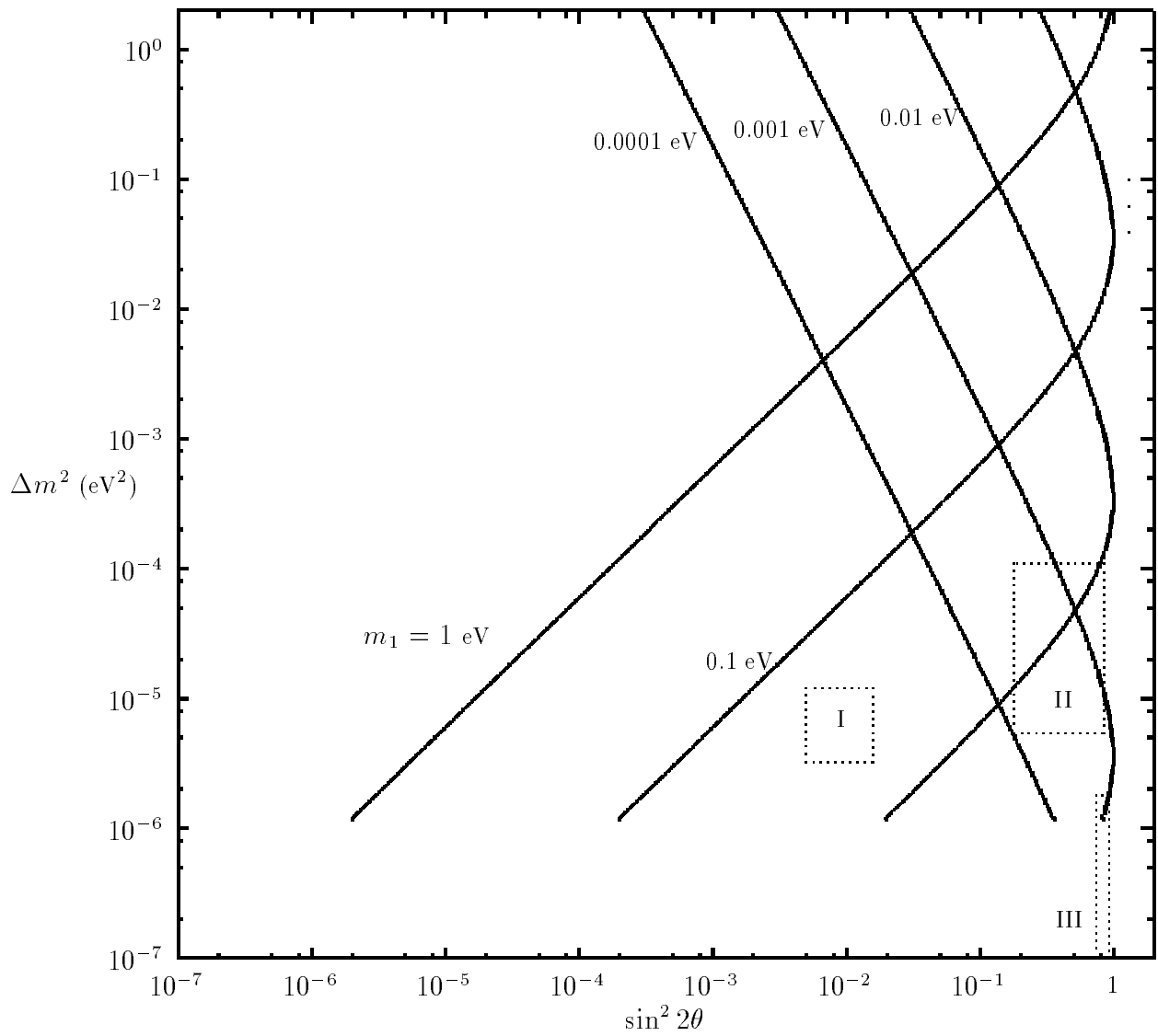
\begin{figure}
\caption{
Plot of the eq.(11) on the parameter space of
$\Delta m^2$ and sin$^2 2\theta$.
It is calculated with the parameter $m_1=$ 1 eV
, 0.1 eV , 0.01 eV , 0.001 eV and 0.0001 eV.
The dashed line boxes denote
the windows permitted by the solar neutrino experiments
coupled to the MSW solution.
}
\label{figone}
\end{figure}
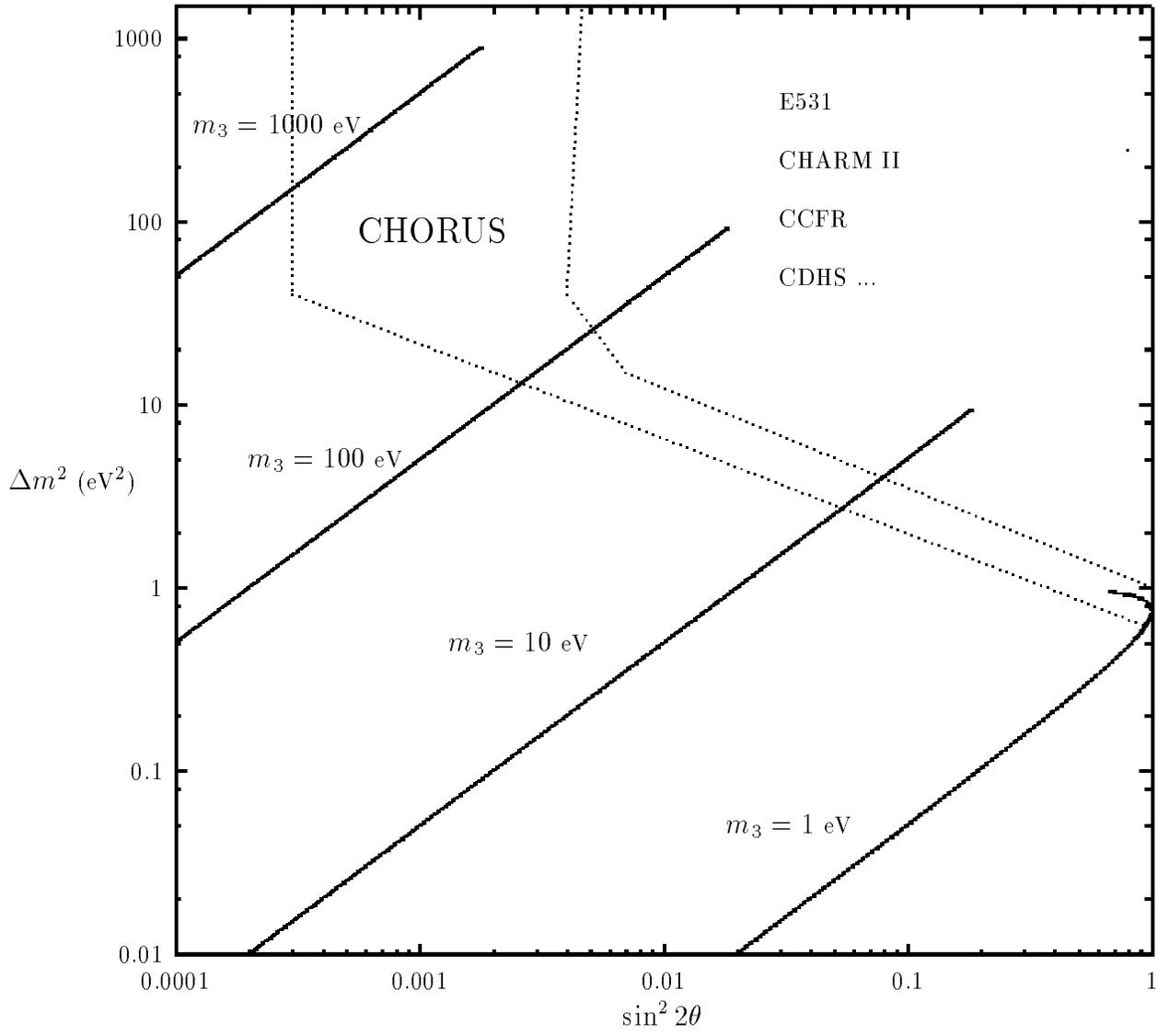
\begin{figure}
\caption{
The predictions of approximate flavour symmetry for
$\nu_{\mu} \leftrightarrow \nu_{\tau}$ oscillation
and the region covered by CHORUS experiment
on the parameter space of
$\Delta m^2$ and sin$^2 2\theta$.
It is calculated with the parameter
$m_3=$ 1 eV, 10 eV, 100 eV and 1000 eV.
}
\label{figtwo}
\end{figure}

\end{document}